\documentclass[12pt]{article}

\usepackage{graphicx}

\usepackage{amsmath}

\usepackage{enumerate}
\usepackage{graphicx}
\usepackage{dcolumn}
\usepackage{bm}
\usepackage{amsmath}
\usepackage{amsxtra}
\usepackage{amstext}
\usepackage{amssymb}
\usepackage{latexsym}
\usepackage{color}         

\usepackage{amscd,verbatim}
\usepackage[all]{xy}

\providecommand{\U}[1]{\protect\rule{.1in}{.1in}}

\begin{document}

\title{Elliptic-Curves Cryptography on High-Dimensional Surfaces}

\author{{Alberto Sonnino$^1$ and Giorgio Sonnino}$^{2,3}$\\
(1) Department of Computer Sciences,\\ University College London (UCL), London, UK\\
Email: alberto.sonnino.15@ucl.ac.uk\\
(2) Department of Theoretical Physics and Mathematics, \\ Universit{\'e} Libre de Bruxelles (U.L.B.), \\
Campus Plaine C.P. 231, 1050  Brussels - Belgium\\
(3) Royal Military School (RMS), \\ Av. de la Renaissance 30, 1000 Brussels - Belgium\\
Email: gsonnino@ulb.ac.be
       }
\date{}
\maketitle
\begin{abstract}
\noindent
We discuss the use of elliptic curves in cryptography on high-dimensional surfaces. In particular, instead of a Diffie-Hellman key exchange protocol written in the form of a bi-dimensional row, where the elements are made up with 256 bits, we propose a key exchange protocol given in a matrix form, with four independent entries each of them constructed with 64 bits. Apart from the great advantage of significantly reducing the number of used bits, this methodology appears to be immune to attacks of the style of Western, Miller, and Adleman, and at the same time it is also able to reach the same level of security as the cryptographic system presently obtained by the Microsoft Digital Rights Management. A nonlinear differential equation (NDE) admitting the elliptic curves as a special case is also proposed. The study of the class of solutions of this NDE is in progress.
\vskip0.5truecm
\medskip\noindent{Keywords: }
{Elliptic-curve cryptography, Elliptic-curve discrete log problem, Public key cryptography, Nonlinear differential equations.}

\vskip1.5truecm
\section{Introduction}

As known, encryption is the conversion of electronic data into another form, called {\it ciphertext}, which cannot be easily understood by anyone except authorized parties. The primary purpose of encryption is to protect the confidentiality of digital data stored on computer systems or transmitted via the Internet or other computer networks. Encryption algorithms can provide not only confidentiality, but also authentication (i.e., the origin message is verified), integrity (i.e., the contents of the message have not been changed), and non-repudiation (i.e., the sender cannot deny to be the author of the message) \cite{ross_anderson, WhatIs}. Elliptic curves are more an more used in cryptography \cite{miller,kapoor}. Their main advantage is that shorter encryption keys use fewer memory and CPU resources for achieving the sam level of security than traditional methods \cite{lauter} . The main concept behind this is the use of the so-called {\it one-way functions}. A one-way function is a function for which it is relatively easy to compute the image of some elements in the domain but it is extremely difficult to reverse this process and determine the original element solely based on the given image \cite{impagliazzo}. More precisely, according to the Federal Office for Information Security (BSI)  \cite{bsi}, the {\it recommended security parameters for elliptic curves is 256 bits} (standards for the years 2017-2021). However, to manipulate data with this degree of security is computationally expensive and often impossible on embedded systems. At present many industrial systems adopt (much) less secure methodologies. This necessitates a re-evaluation of our cryptographic strategy. The question is: {\it are we able to obtain the same degree of security with small embedded microprocessors managing only 64-bit operations}? The solution of this problem entails several steps:

\noindent {\bf A)} First step: {\it Research}

\noindent The solution of this problem requires new mathematical concepts and algorithms. 

\noindent {\bf B)} Second step: {\it Commercialization}

\noindent Once found the solution, the process is concluded with the start-up of the commercialization of the product.

\noindent This manuscript deals only with the first step. We shall introduce a hyper-surface in an arbitrary $(n^2+1)$-dimensional space (with $n$ denoting a positive integer number), and we use the idea of the {\it one-way function} possessing also the property of being a {\it trap function}. The encrypted shared-key, instead to be written as a (very large) scalar number is brought into a matrix form. We shall prove that we may obtain the same degree of security as the one obtained by the Microsoft Digital Rights Management  \cite{krawczyk} by sending an encrypted shared-matrix with four independent entries, each of them made up by 64 bits. The encrypted information is successively transmitted through elliptic curves obtained by projecting the hyper-surface imbedded in a $(n^2+1)$-dimensional space onto perpendicular planes. This methodology allows reaching the same level of security as the cryptographic system presently obtained by the Microsoft Digital Rights Management.

\noindent The manuscript is organized as follows. In Section~(\ref{cryptography}) we introduce high-dimensional surfaces cryptography (HDSC) and the elliptic curves constructed through these hyper-surfaces. Without loss of generality, we shall limit ourselves to the case of $n=2$ (i.e., to a $5D$-space). The generalization to $(n^2+1)$-dimensional space is straightforward. The definition of the groups in elliptic curves on high-dimensional surfaces and the elliptic curve discrete log problem can be found in the Subsections~(\ref{groups}) and ~(\ref{hdlogproblem}), respectively. Concluding remarks are reported in the Section~(\ref{conclusions}).

\section{Elliptic-Curves Cryptography on High Dimensional Surfaces}\label{cryptography}

We illustrate the methodology by dealing with a five-dimensional elliptic curve, even though the procedure is straightforwardly generalizable to elliptic curves on surfaces imbedded in an arbitrary $(n^2+1)$-dimensional space, with $n$ denoting a positive integer number (the reason for which only spaces of such dimension are allowed will soon be clear). For the sake of simplicity, in this work we shall limit ourselves to the analysis of elliptic curves on 5-dimensional surfaces. The generalization to the general case (i.e., to the case of elliptic curves on $(n^2+1)$-hyper-surface) is straightforward\footnote{In fact, we anticipate that the encrypted code involves only square matrices of order $n\times n$  - see next Section.}. In a 5-dimensional space, the surfaces on which the elliptic curves are defined, are the solutions of the equation 
\begin{align}\label{hdecc1}
&E=\Big\{(y,x_1,x_2,x_3,x_4)\mid y^2=x_1^3+x_2^3+x_3^3+x_4^3+{\mathbf A}\cdot{\mathbf X}+b\Bigr\}\qquad {\rm where}\\
&{\mathbf A}\equiv\begin{pmatrix}
a_{1}, & a_{2}, & a_{3}, & a_{4}
\end{pmatrix}
\qquad {\rm ;} \qquad
{\mathbf X}\equiv
\begin{pmatrix}
x_1\\
x_2\\
x_3\\
x_4
\end{pmatrix}\nonumber
\end{align}
\noindent with $a_i$ ($i=1,\cdots , 4$) and $b$ denoting elements of the field $\mathbb{K}$. Examples of fields $\mathbb{K}$ are the Real Numbers, $\mathbb{R}$, the Rational Numbers, $\mathbb{Q}$, the Complex Numbers, $\mathbb{C}$ or the Integers modulo $p$, $\mathbb{Z}/p{\mathbb Z}$. By fixing three of the four variables $x_i$ (by setting, for example, $x_2=c_2=const.$, $x_3=c_3=const.$ and $x_4=c_4=const.$) and by rotating the indexes, Eq.~(\ref{hdecc1}) defines together with the {\it points at infinity} $\mathcal{O}$, four distinguished two-dimensional elliptic curves $E_i$:
\begin{align}\label{hdecc2}
&E_i=\Big\{(y,x_i)\mid y^2=x_i^3+a_ix_i+b_i\Bigr\}\cup\bigl\{\mathcal O\bigr\}\quad {\rm with}\quad i=(1,\cdots,4)\quad {\rm and}\\
&b_1\equiv b+c_2^3+c_3^3+c_4^3+a_2c_2+a_3c_3+a_4c_4\nonumber\\
&b_2\equiv b+c_1^3+c_3^3+c_4^3+a_1c_1+a_3c_3+a_4c_4\nonumber\\
&b_3\equiv b+c_1^3+c_2^3+c_4^3+a_1c_1+a_2c_2+a_4c_4\nonumber\\
&b_4\equiv b+c_1^3+c_2^3+c_3^3+a_1c_1+a_2c_2+a_3c_3\nonumber
\end{align}
\noindent being $c_i$ ($i=1\cdots 4$) elements of $\mathbb{K}$. In order to avoid degeneracy, these parameters are subject to the following restrictions
\begin{align}\label{hdecc3}
&4a_1^3+27(b+c_2^3+c_3^3+c_4^3+a_2c_2+a_3c_3+a_4c_4)^2\neq 0\\
&4a_2^3+27(b+c_1^3+c_3^3+c_4^3+a_1c_1+a_3c_3+a_4c_4)^2\neq 0\nonumber\\
&4a_3^3+27(b+c_1^3+c_2^3+c_4^3+a_1c_1+a_2c_2+a_4c_4)^2\neq 0\nonumber\\
&4a_4^3+27(b+c_1^3+c_2^3+c_3^3+a_1c_1+a_2c_2+a_3c_3)^2\neq 0\nonumber
\end{align}
\noindent Clearly, in case of $\mathbb{K}=\mathbb{Z}/p\mathbb{Z}$, we may associate four modules $p$ to each elliptic curves. As we shall see in the forthcoming section, only elliptic curves on hyper-surfaces of dimension $n^2 + 1$ (with $n=$ denoting a positive integer number) are acceptable since the shared-key involves only square matrices of order $n\times n$. For illustration purpose only, Figure~\ref{curve_3D} shows a three dimensional surface where the values of the parameters are $a_1=-4$, $a_2=-5$ and $b=3.5$. Figures~\ref{curve_fixed_x1} and \ref{curve_fixed_x2} refer to the elliptic curves obtained by projecting the surface~\ref{curve_3D} onto the planes $x_1=1$ and $x_2=-2$, respectively. 
\begin{figure*}[htb] 
\hspace{4cm}\includegraphics[width=8cm,height=8cm]{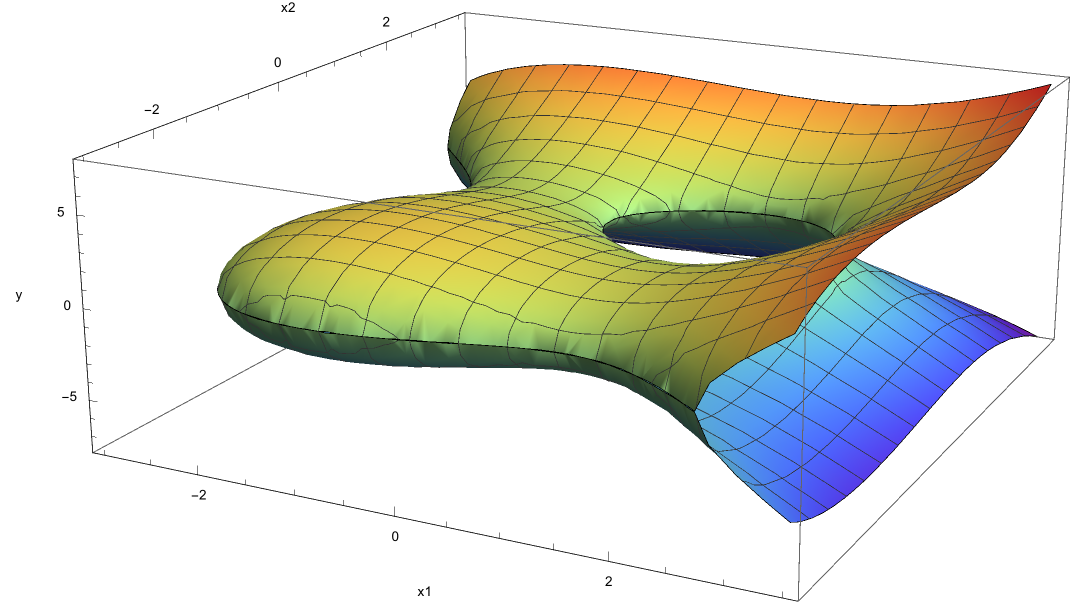}
\caption{ \label{curve_3D} Only for illustration purpose, we show the two-dimensional surface given by Eq.~(\ref{hdecc2}) with parameters $a_1=-4$, $a_2=-5$ and $b=3.5$. However, one should bear in mind that only elliptic curves on hyper-surfaces of dimension $n^2 + 1$ have real meaning (hence, only elliptic curves constructed by hyper-surfaces of dimensions $2$, $5$, $10$ and so on, are acceptable). This because, as we shall see in the forthcoming section, the encrypted code involves only square matrices of order $n\times n$.}
\end{figure*}
\begin{figure*}[htb] 
\hfill 
\begin{minipage}[t]{.45\textwidth}
    \begin{center}  
\hspace{-1.2cm}
\resizebox{1\textwidth}{!}{%
\includegraphics{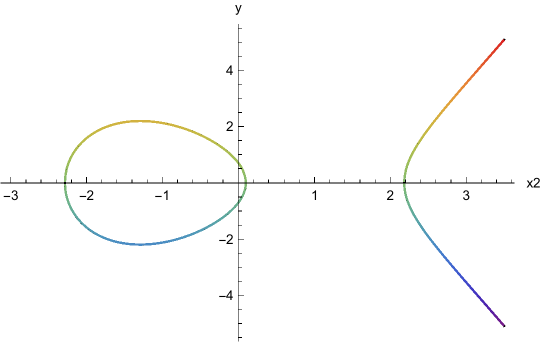}
}
\caption{ \label{curve_fixed_x1}Elliptic curve obtained by projecting the surface~\ref{curve_3D} onto the plane $x_1=1$.}
\end{center}
  \end{minipage}
\hfill 
\begin{minipage}[t]{0.5\textwidth}
    \begin{center}
\hspace{-0.9cm}
\resizebox{1\textwidth}{!}{%
\includegraphics{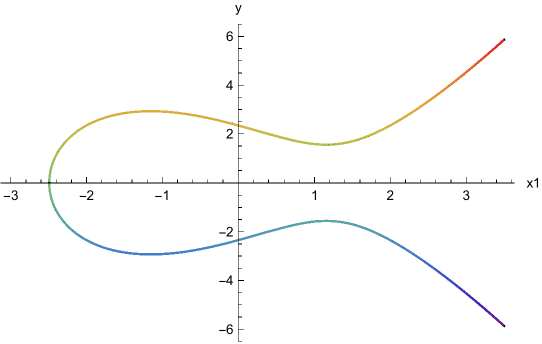}
}
\caption{Elliptic curve obtained by projecting the surface~\ref{curve_3D} onto the plane $x_2=-2$.}
\label{curve_fixed_x2}
\end{center}
  \end{minipage}
\hfill
\end{figure*}
\subsection{Groups in Elliptic Curves on High-Dimensional Surfaces}\label{groups}
Each elliptic curve $E_i$, separately, defines under point addition an abelian group. For each $P_i\in E_i$, $Q_i\in E_i$ and $R_i\in E_i$ the following properties are satisfied:

\noindent$\bullet$ {\it Commutative} : $P_i+Q_i=Q_i+P_i$.

\noindent$\bullet$  {\it Identity} : $P_i+\mathcal O=\mathcal O +P_i=P_i$;

\noindent$\bullet$ {\it Inverse} : $P_i-P_i=P_i+(-P_i)=\mathcal O$;

\noindent$\bullet$ {\it Associative} : $P_i+(Q_i+R_i)=(P_i+Q_i)+R_i$;

\noindent $\bullet$ {\it Closed} :  If $P_i\in E_i$ and $Q_i\in E_i$, then $P_i+Q_i\in E_i$;

\vskip0.5truecm

\noindent Each group identified by $E_i$ is equipped with the standard {\it group operations} \cite{Weisstein, blake}, i.e.

\noindent$\bullet$ {\it Addition} - If $P_i=(P_{ix_i},P_{iy})\in E_i$ and $Q_i=(Q_{ix_i},Q_{iy})\in E_i$, then $P_i+Q_i=R_i$ [with $R_i=(R_{ix_i},R_{iy})$], which algebraically is defined as
\begin{align}\label{hdecc4}
&R_{ix_i}=s_i^2-(P_{ix_i}+Q_{ix_i})\\
&R_{iy}=s_i(P_{ix_i}-R_{ix_i})-P_{iy}\nonumber\\
&s_i=\frac{P_{iy}-Q_{iy}}{P_{ix_i}-Q_{ix_i}}\nonumber
\end{align}
\noindent for $i=(1,\cdots,4)$. As a particular case, we get $2P$: 
\begin{align}\label{hdecc5}
&2P_{ix_i}=s_i^2-2P_{ix_i}\\
&2P_{iy}=s_i(P_{ix_i}-R_{ix_i})-P_{iy}\nonumber\\
&s_i=\frac{3P_{ix_i}^2+a_i}{2P_{iy}}\nonumber
\end{align} with $i=(1,\cdots,4)$. The points at infinity are reached in each elliptic curves $E_i$ when $P_i+Q_i=\mathcal O$ if $P_{ix_i}=Q_{ix_i}$ or when $y=0$ for point doubling (i.e., $P_i+P_i=\mathcal O$).

\noindent$\bullet$ {\it Scalar Multiplication} - If $P\in E_i$ and $\kappa\in\mathbb{Z}$, Eq.~(\ref{hdecc5}) allows defining the operation $Q=\kappa P$ under the condition that the operation $Q=\kappa P\equiv P+\cdots +P$, equal $\kappa$ times $P$, is performed by using the {\it same} elliptic curve $E_i$ i.e., $Q\in E_i$. The scalar multiplication defines the {\it one-way function} $Q\rightarrow P$ where is very difficult to extract the value of $\kappa$.

\noindent $\bullet$ {\it Reflection} - The reflection of a point is its inverse. Hence for $P_i=(P_{ix_i},P_{iy}$) the inverse of $P_i$ is $-P_i=(P_{ix_i},-P_{iy}$) satisfying the relation $P_i-P_i=\mathcal O$.

\subsection{High-Dimensional Elliptic Curve Discrete Log Problem}\label{hdlogproblem}
For each $E_i$ the scalar multiplication defines a {\it one way-function} \cite{menezes}. Let us consider elliptic curves $E_i(\mathbb{Z}/p_i\mathbb{Z}$), with $p_i=(p_1,\cdots, p_4)$, and let and et $Q_1$ and $P_1$ two points belonging to the same elliptic curve, say $E_1$, with the condition that $Q_1$ is a multiple of $P_1$. As know, to find the value of the number $\kappa$ such that $Q_1=\kappa P_1$ is a very difficult problem \cite{smart}. We introduce now the first {\it base point} ({\it Generator}), $G_1\equiv (G_{x_1}, G_y)\in E_1({\mathbb Z}/p_1\mathbb{Z})$. Since the group is {\it closed}, $G_1$ generates a {\it cyclic group} under point addition in the curve $E_1$. The order $n_1$ (with $n_1\in\mathbb{K}$) of $G_1$ is the number of the points in the group that $G_1$ generates. By this operation, we say that $G_1$ generates a subgroup of size $n$, and we write $ord(G_1)=n_1$. The {\it order of the subgroup} generated by $G_1$ is the smallest integer $\kappa_1$ such that $\kappa_1G_1=\mathcal O$ (hence, $n_1<\kappa_1$). 

\noindent After $n_1$ iterations on the curve $E_1$ we find a second {\it base point} ({\it Generator}), $G_2$ with coordinates $G_2(n_1)=[G_{x_2}(n_1),G_{y_2}(n_1)]$. We may keep this second generator to perform $n_2$ iterations on the curve $E_2$, with $n_2<\kappa_2$ being $\kappa_2$ the order of the subgroup generated by $G_2$ on the elliptic curve $E_2$. After $n_2$ iterations we get a third {\it base point} ({\it Generator}), $G_3$ with coordinates $G_3(n_2)=[G_{x_3}(n_2),G_{y_3}(n_2)]$. With this second generator we perform $n_3$ iterations on the curve $E_3$, with $n_3<\kappa_3$ (with $\kappa_"$ denoting the order of the subgroup generated by $G_3$ on the elliptic curve $E_3$). After $n_3$ iterations on the curve $E_3$ we get the fourth {\it base point} ({\it Generator}), $G_4$ with coordinates $G_4(n_3)=[G_{x_4}(n_3),G_{y_4}(n_3)]$. The process concludes after $n_4$ iterations on the elliptic curves $E_4$ (with $n_4$ less than $\kappa_4$, the order of the subgroup generated by $G_3$ on the curve $E_4$). At the end of these operations we get three matrices $N$, $G$ and $K$, of order $2\times 2$, where the entries are totally {\it independent from each others}. Matrices $N$ and $G$ reads\footnote{Note that, once generated, the elements $G_{x_1}, G_{x_i}(n_i)$ may be allocated as entries of the matrix $G$ in a random way.}
\begin{equation}\label{hdecc6}
N=
\begin{pmatrix}
n_1 & n_2 \\
n_3& n_4\
\end{pmatrix}\quad ;\quad 
G=
\begin{pmatrix}
G_{x_1} & G_{x_2}(n_1) \\
G_{x_3}(n_2)& G_{x_4}(n_3)\
\end{pmatrix}\quad ;\quad
K=\begin{pmatrix}
\kappa_1 & \kappa_2 \\
\kappa_3& \kappa_4\
\end{pmatrix}
\end{equation}
\noindent The parameters that also {\it Eve}, the eavesdropper, possesses are ($p_i, a_i, b_i, G, K$) with $i=(1,\cdots, 4)$. $p_i$ specify the modulo of the fields $\mathbb{K}_i$, $a_i$ and $b_i$ define the elliptic curves $E_i$ (notice that in general these curves are different from each others), $G$ is the Generator matrix and $K$ is the order of the subgroups generated by $G$, respectively. Now, if {\it Bob} and {\it Alice} want to communicate with each other, Bob picks private key $N$ with $1\leq n_i\leq\kappa_i-1$, $i=(1,\cdots, 4)$. Bob computes matrix $T=NG$, which belongs to the curves $E$ [given by Eq.~(\ref{hdecc1})]. At the same time, Alice picks private key $M$ 
\begin{equation}\label{hdecc7}
M=\begin{pmatrix}
m_1 & m_2 \\
m_3& m_4\
\end{pmatrix}
\end{equation}
\noindent where the entries $m_i$ satisfy the conditions $1\leq m_i\leq\kappa_i-1$, $i=(1,\cdots, 4)$. {\it Alice} receives from {\it Bob} the information $T$ and she generates the point $MT=MGN=W$ (notice that matrices do not commute). {\it Bob} receives from {\it Alice} the information $P=MG$ and he computes $PN=MGN=W$\footnote{Note that {\it Bob} multiplies matrices by placing $N$ always on the right, while {\it Alice} multiplies matrices by placing $M$ always on the left.}. Both players, {\it Bob} and {\it Alice}, possess the same (encrypted) key $W$, which also belongs to the curve $E$ [given by Eq.~(\ref{hdecc1})]. {\it Eve}, the eavesdropper, sees both information $T$ and $P$, but he is unable to retrieve the sheared-key $W$. Figure~\ref{scheme_hdecc} depicts the entire process.
\begin{figure*}[htb] 
\hspace{4cm}\includegraphics[width=8cm,height=8cm]{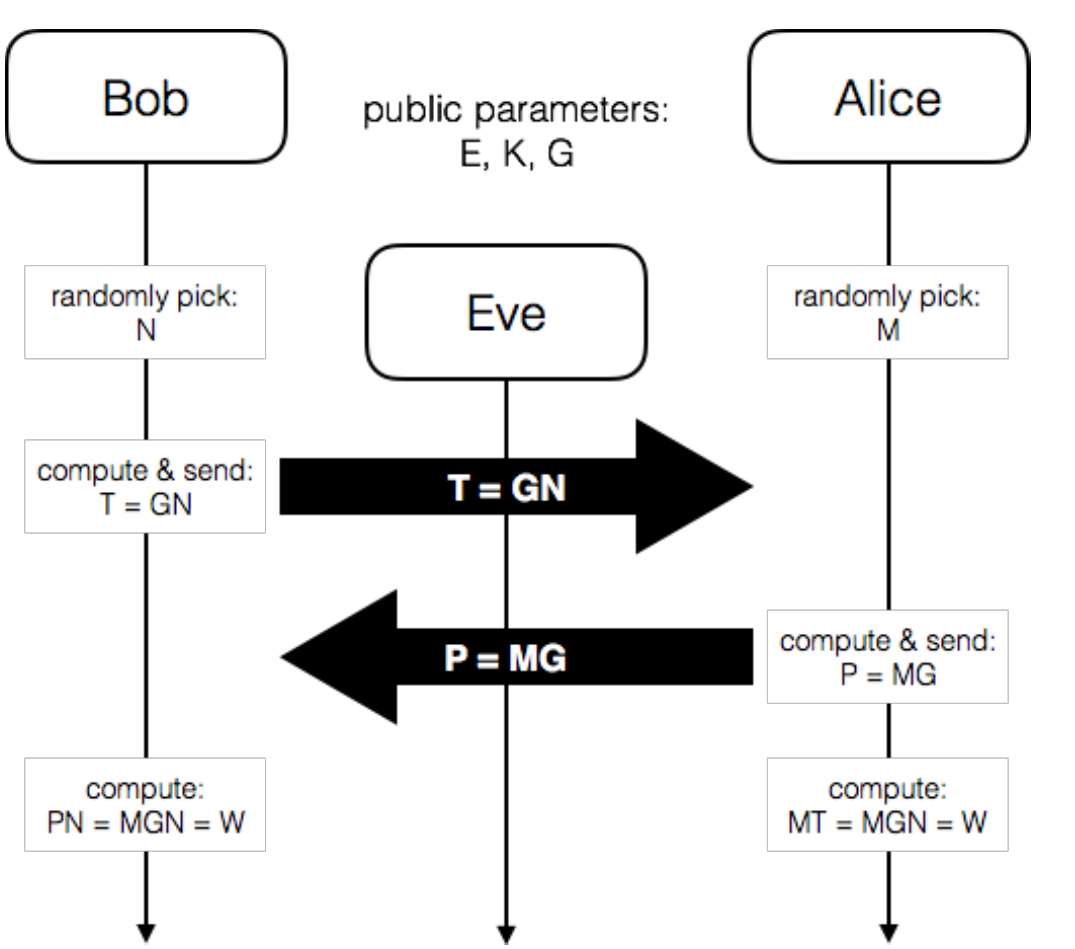}
\caption{ \label{scheme_hdecc} Diffie-Hellman key exchange protocol \cite{diffie_hellman} in high-dimensional elliptic curves cryptography. {\it Bob} and {\it Alice} exchange the encryption key in a matrix form, $W=MGN$, having four independent entries each of them constructed with 64 bits. {\it Eve}, the eavesdropper, may see $T=GN$ and $P=MG$, but he is unable to decrypt the sheared {\it Bob}-{\it Alice}'s key $W$ since it is very difficult to reverse the process and determine what was the original information.}
\end{figure*}

\noindent$\bullet$ {\it Eve} does not know to which entry of the matrix G the generators have been assigned;

\noindent$\bullet$ In case of $5D$-elliptic curves, the process runs on four, distinguished and independent, $2D$-elliptic curves and the encrypted key belongs to a $5D$-surface. This hyper-surface is constructed in such a way that the curves with variables $x_i$, obtained by setting constant the remaining variables of this hyper-surface (i.e., by setting $x_j=const.$ with $j\neq i$), are elliptic curves;

\noindent $\bullet$ The level of security remains unchanged. Indeed, it is easily to convince ourselves that to obtain the same level of security as in case of one-dimensional elliptic curve cryptography (which requires 256 bits), we need to encode the shared-key with only 64 bits elements (since in our case, for a shared-key written in the form of a matrix $2\times 2$ the level of security is of the order of $\alpha^4$, with $\alpha$ denoting the number of required bits).

\noindent We recall that the present methodology applies to elliptic curves cryptography constructed on hyper-surfaces of dimension $n^2+1$ (with $n$ denoting an integer number) because the shared-key is brought into the form of $n\times n$ square matrices. Hence, a surface like Eq.~(\ref{hdecc1}) is the immediate generalization of a one-dimensional elliptic curves cryptography. The subsequent surface which generalizes Eq.~(\ref{hdecc1}) should be imbedded in a ten-dimensional space, and so on.

\section{Examples of Practical Uses of High- Dimensional Elliptic Curve Cryptography }\label{examples}
The aim of this section is to illustrate the many possibilities and practical usages opened by the introduction of High-Dimensional Elliptic Curve Cryptography. Generally speaking, HDECC can be applied instead of any use of the classic ECC; i.e., Bitcoin, secure shell ({\it ssl}), transport layer security ({\it tls}) \cite{bos}.  Among these applications, one of the most important is certainly {\it tls}. Indeed, {\it tls} is the new generation of the Secure Socket Layer ({\it ssl}) which is used in any modern telecommunication. For instance, the well-known {\it https} is nothing else than the classic {\it http} protocol running within {\it ssl/tls} in order to ensure a secured, bidirectional connection for arbitrary binary data between two hosts. In order to establish a shared key between these two hosts, the current implementations of {\it tls} mainly relies on the DH or ECDH key exchange protocols discussed in the previous sections.

\noindent
However, introducing such security layers comes at the price of overheads in terms of infrastructure costs, communication latency, data usage, and energy consumption \cite{naylor}. Therefore, the first motivation of HDECC is to reduce the cost of security in many of the today's state-of-the-art communication technologies. Moreover, reducing these costs makes the most modern security protocols accessible for embedded systems and wearable devices. Indeed, by using HDECC, we could reduce the cost of these protocols by performing operations on data four time shorter than before by maintaining, at the same time, the same level of security. In addition, HDECC opens new perspectives on elliptic curve cryptography as we shall discuss in the next section.
\section{Conclusion and Perspectives}\label{conclusions}

We have proposed an encrypted procedure based on the high-dimensional elliptic curve cryptography, which  allows maintaining the same level of security as presently obtained by the Microsoft Digital Rights Management. The advantages of these methodology are multiplex.         

\noindent $\bf 1)$ The quite heavy intermediate exponential operations are avoided and the key exchange protocol is constructed with 64 bits operations instead of 256 bits.

\noindent $\bf 2)$ We may proceed to the construct of a new generation of cryptographic standards working with the technology {\it high-dimensional elliptic curves}. 

\noindent $\bf 3)$ This methodology opens new perspectives. In fact it is not difficult to derive a nonlinear differential equation (NDE) admitting the elliptic curves by a special choice of the parameters and initial conditions. We get
\begin{align}\label{c1}
&y^{\prime\prime}+\alpha_1(x)y^{-1/2}y^{\prime\prime}+\alpha_2(x)y^{-1/2}y^\prime+\alpha_3(x)y^{-3/2}y^{{\prime}^2}+\alpha_4(x)=0\\
&y(0)=\beta_1\quad ;\quad y^\prime (0)=\beta_2\nonumber\\
&{\rm with}\quad \alpha_i(x)=a_i+b_ix\quad (i=1,2,3,4)\quad {\rm and}\quad a_i,\ b_i,\ \beta_1,\ \beta_2=const.\nonumber
\end{align}
\noindent with ${}^{\prime}$ denoting the derivative with respect to the variable $x$. Note that the differential equation~(\ref{c1}) admits as a special solution the {\it Weierstrass equation} \cite{laska}
\begin{equation}\label{c2}
y^2+c_1xy+c_3y=x^3+c_2x^2+c_4x+c _6
\end{equation}
\noindent with $c_i\in K$. If $Char K\neq (2,3)$, we can complete before the square and, successively, the cube, by defining
\begin{equation}\label{c3}
\eta=y+(c_1x+c_3)/2\qquad ;\qquad \xi=x+(c_1^2+4c_2)/12
\end{equation}
\noindent By substituting Eqs~(\ref{c3}) into Eq.~(\ref{c2}), we get the {\it elliptic curve} \cite{connell}:
\begin{equation}\label{c4}
\eta^2=\xi^3-\frac{d_4}{48}\xi-\frac{d_6}{864}
\end{equation}
\noindent where
\begin{align}\label{c5}
&d_4= (c_1^2+4c_2)^2-24(c_1c_3+2c_4)\nonumber\\
&d_6= -(c_1^2+4c_2)^3+36(c_1^2+4c_2)(c_1c_3+2c_4)-216(c_3^2+4c_6)
\end{align}
\noindent Clearly, now the question is: {\it how can we determine the largest class of parameters $a_ i$, $b_i$, $\beta_1$ and $\beta_2$, introduced in (\ref{c1}), such that the NDE~(\ref{c1}) admits (only) a class of one-way functions, possessing the property of being trap functions} ? In addition, {\it we should also be able to define on these curves a group under point addition}. Successively, the trapped curves could be identified uniquely by indexes. Being able to answer to this question would allow encrypting not only the key exchange protocol but also the trapped-curves on which the generator and the encrypted keys belong. However, all of this requires sophisticated mathematical tools and it will be subject of future works.

\noindent We close this Section by mentioning other two relevant perspectives of this work. 

\noindent {\bf i)} It is quite evident that the formalism illustrated in this manuscript allows introducing two operations: {\it matrix addition} and {\it scalar matrix multiplication} (including the so-called {\it matrix doubling operation}). These operations can be used to implement a high-dimensional version of algorithms such as the ECDSA (elliptic curves digital signature algorithm) \cite{ecdsa}.

\noindent {\bf ii)} It is possible to introduce an operator ${\mathcal L}$ which connects two distinct points $G^{(1)}$, $G^{(2)}$ on the high-dimensional surface $E$ [see Eq.~(\ref{hdecc1})] as follows
\begin{equation}\label{c6}
G^{(1)}={\mathcal L} G^{(2)}\quad\Longrightarrow \quad 
\begin{pmatrix}
G^{(1)}_{x_1} & G^{(1)}_{x_2}(n_1) \\
G^{(1)}_{x_3}(n_2)& G^{(1)}_{x_4}(n_3)\
\end{pmatrix}
={\mathcal L}
\begin{pmatrix}
G^{(2)}_{x_1} & G^{(2)}_{x_2}(n_4) \\
G^{(2)}_{x_3}(n_5)& G^{(2)}_{x_4}(n_6)\
\end{pmatrix}
\end{equation}
\noindent with $\mathcal L$ denoting a non-singular $2\times 2$ matrix, satisfying the group law under matrix multiplication. The analytic expression and the mathematical study of this matrix (and the $n\times n$ matrices, in general), with its potential application in cryptography, will be subject of a future work.

\section{Acknolwedgments}\label{Ack}
AS is indebted with Prof. G. Danezis, from University College London (UCL), Department of Computer Sciences, and Prof. J. Becker, from Karlsruhe Institute of Technology (KIT), Institut f${\ddot{\rm u}}$r Technik der Informationsverarbeitung (ITIV), for their support and useful suggestions. GS is also very grateful to Prof. Pasquale Nardone and Dr Philippe Peeters from the Universit{\`e} Libre de Bruxelles (U.L.B.). 


\vskip1truecm


\end{abstract}
\thispagestyle{empty}
\end{document}